\documentclass[a4paper,11pt]{article}
\pdfoutput=1 % if your are submitting a pdflatex (i.e. if you have
\usepackage{jcappub}
\usepackage{aas_macros}
\usepackage[T1]{fontenc}

\usepackage[utf8]{inputenc} % allow utf-8 input
\usepackage[T1]{fontenc}    % use 8-bit T1 fonts
\usepackage{hyperref}       % hyperlinks
\usepackage{url}            % simple URL typesetting
\usepackage{booktabs}       % professional-quality tables
\usepackage{amsfonts}       % blackboard math symbols
\usepackage{nicefrac}       % compact symbols for 1/2, etc.
\usepackage{microtype}      % microtypography

\title{Inferring galaxy cluster masses from cosmic microwave background lensing with neural simulation based inference}

\author[1]{Eric J. Baxter,\note{Corresponding author.}}
\author[2]{Shivam Pandey}

\affiliation[1]{Institute for Astronomy, University of Hawai`i, 2680 Woodlawn Drive, Honolulu, HI 96822, USA}
\affiliation[2]{Columbia Astrophysics Laboratory, Columbia University, 550 West 120th Street, New York, NY 10027, USA}

\emailAdd{ebax@hawaii.edu}
\emailAdd{sp4204@columbia.edu}

\abstract{
    Gravitational lensing by massive galaxy clusters distorts the observed cosmic microwave background (CMB) on arcminute scales, and these distortions carry information about cluster masses.  Standard approaches to extracting cluster mass constraints from the CMB cluster lensing signal are either sub-optimal, ignore important physical or observational effects, are computationally intractable, or require additional work to turn the lensing measurements into constraints on cluster masses.  We apply simulation based inference (SBI) using neural likelihood models to the problem.  We show that in circumstances where the exact likelihood can be computed, the SBI constraints on cluster masses are in agreement with the exact likelihood, demonstrating that the SBI constraints are close to optimal.  In scenarios where the exact likelihood cannot be feasibly computed, SBI still recovers unbiased estimates of individual cluster masses and combined constraints from multiple clusters.  SBI will be a powerful tool for constraining the masses of galaxy clusters detected by future cosmic surveys.  Code to run the analyses presented here will be made publicly available.
}
\begin{document}
\maketitle
\flushbottom

\section{Introduction}

The abundance of galaxy clusters is highly sensitive to cosmological parameters such as $\sigma_8$ and $\Omega_{\rm m}$, as well as to possible departures from general relativity on cosmological scales \citep{DETF}.  However, using the abundance of galaxy clusters as a cosmological probe requires accurate calibration of cluster masses: the cluster abundance falls exponentially with increasing mass, so small errors in cluster mass measurements can lead to significant biases in the inferred cosmological parameters.  Future cosmic microwave background (CMB) surveys like CMB Stage 4 (CMB-S4; \cite{S4}) will detect of order $10^5$ clusters at high redshift ($z \gtrsim 1$) via the Sunyaev Zel'dovich (SZ) effect \cite{Raghunathan:2022}, caused by CMB photons inverse Compton scattering with hot gas inside of clusters \cite{SZ}.  Fully harnessing the statistical power of these future cluster samples --- more than an order of magnitude larger than current samples \citep[e.g][]{Bleem:2015, Bleem:2020, Hilton:2021} --- will necessitate high-precision constraints on cluster masses over a broad redshift range.  Gravitational lensing provides a powerful way to obtain these constraints because it is sensitive to all mass in the cluster, and does not rely on assumptions about e.g. the thermal state of gas within the cluster.  Because the signal-to-noise per cluster is typically low for lensing measurements, one often uses the lensing measurements primarily to calibrate the overall normalization of the relationship between the  cluster mass and some cluster observable which can be measured at high signal-to-noise, like the cluster SZ or X-ray signal.

However, measuring the lensing signal from the distortions of galaxy shapes is not feasible for high redshift clusters because of the difficulty of obtaining galaxy shape measurements for high redshift background galaxies.  Moreover, galaxy lensing is subject to a number of difficult to model sources of systematic uncertainty, including photometric redshift errors, boost factors and intrinsic alignments \citep[e.g.][]{McClintock:2019}.  

Another powerful way to calibrate the masses of galaxy clusters at high redshift is to instead use gravitational lensing of the CMB.  Gravitational lensing by galaxy clusters causes small distortions in the CMB on arcminute scales \citep{Seljak:2000}.  These distortions can be estimated, and used to constrain the masses of the lensing clusters.  CMB lensing is not impacted by intrinsic alignments, photometric redshift errors or boost factors, making it a powerful crosscheck of galaxy lensing measurements.  Moreover, because the CMB originates at $z\sim 1100$, CMB cluster lensing can be used to calibrate the masses of clusters at essentially arbitrarily high redshift.

The standard approach to measuring the CMB lensing signal is to use estimators formed from quadratic combinations of the Fourier modes of the observed CMB field at different multipoles.  In the absence of lensing, the expectation value of these quadratic terms vanishes; in the presence of lensing, an appropriately weighted combination of these quadratic terms will provide an estimate of the lensing deflection field \cite{HuOkamoto:2002}.  The standard quadratic estimator is biased in the presence of strong lensing by the cluster, but this bias can be removed with an acceptable degradation in signal-to-noise by using an appropriate choice of filtering \citep{HuCluster:2007}.  Once the deflection maps (or some related quantity, such as the lensing convergence) have been estimated with a quadratic estimator, they can be fit with parameterized models for the cluster mass distribution to obtain constraints on the cluster masses, which are ultimately the parameters of interest.  This approach has been applied in several analyses \citep[e.g.][]{Madhavacheril:2015, Melin:2015, Baxter:2018, Madhavacheril:2020}.  

However, at small scales, the quadratic estimator approach is known to be  sub-optimal \citep[e.g.][]{Raghunathan:2017, Horowitz:2019, Hadzhiyska:2019} and will lead to unnecessarily weak constraints on the cluster masses by factors of order two at noise levels comparable to those expected for CMB-S4 \citep{Raghunathan:2017, Saha:2023}.  Additionally, since the cluster mass is ultimately the parameter of interest, the intermediate step of generating a lensing deflection or convergence map is not strictly necessary, and can present additional challenges in practice.  For instance, a standard approach is to compresses the lensing measurements to a summary such as the convergence profile, and to then fit the profile measurements to a model which depends on the cluster mass.  This process can result in loss of information, and also requires additional work to estimate the covariance of the convergence profile measurements.\footnote{Of course, there are also advantages to estimating a lensing convergence map.  For instance, such maps provide an easy to interpret summary of the lensing information.}  

An alternative to the quadratic estimator approach is to compute the likelihood of the observed CMB given the known statistics of the unlensed CMB and a model for the cluster mass distribution.  \cite{Dodelson:2004, Baxter:2015, Raghunathan:2017}.  In principle, if this likelihood could be computed, it would yield optimal constraints on the parameters of the cluster mass distribution without the need to construct the deflection field map and perform additional fitting to the cluster profile, as is done in quadratic estimator-based analyses.  However, outside of highly idealized scenarios, computing the exact likelihood for the lensed CMB becomes difficult.  For example, the telescope beam and anisotropic filtering of the observed CMB necessitate many calculations of high-dimensional integrals over the latent space of the possible unlensed and un-beamed CMB realizations \cite{Baxter:2015}.  

Further complicating the likelihood calculation is the fact that the CMB is also lensed by other large scale structure (LSS) along the line of sight in addition to the galaxy cluster.  Some of this structure will be in front of the cluster (i.e. at lower redshift), while some will be behind (i.e. at higher redshift), leading to complex changes to the observed CMB that are difficult to include exactly in a likelihood approach. 

A number of other methods beyond quadratic estimators and the exact likelihood have been considered for extracting the CMB cluster lensing signal.  \cite{Saha:2023} use the techniques of maximum a posteriori (MAP) lensing reconstruction to constrain the lensing profiles of galaxy clusters, taking into account lensing by LSS.  These profiles can then be fit to extract cluster mass constraints.  These constraints, like those of the exact likelihood approach, should be optimal.
\cite{Gupta:2021} demonstrated that a two-stage neural network based on \texttt{ResUNet} \cite{resunet} could be trained to predict the masses of galaxy clusters from observations of the lensed CMB.  \cite{Raghunathan:2019} stack maps of CMB temperature or polarization along directions of the locally measured CMB gradient, and fit the resultant lensing-induced dipole to constrain the cluster masses.

Here, we present a new method for inferring the masses of galaxy clusters from CMB lensing that relies on simulation based inference (SBI).  While analytically computing the exact likelihood becomes complicated outside of highly idealized scenarios, simulating realizations of the cluster-lensed CMB in full complexity is straightforward and fast.  One simply generates a Gaussian realization of the unlensed CMB, remaps this field using a simulated lensing deflection map for the cluster and for additional line-of-sight structures, and then adds whatever observational effects (e.g. beam and filtering) are appropriate.  The key point is that simulation requires only drawing from the high-dimensional distributions of primordial CMB, LSS, and noise realizations, while computing the likelihood requires integrating over these distributions.  The intractability of the likelihood and the ease of simulation make the problem especially well suited to the techniques of simulation based inference (SBI).  In SBI, one uses simulations to obtain a sufficiently accurate approximation to the posterior, rather than computing it exactly.  

We apply several related SBI methods to the problem, focusing on Neural Ratio Estimation (NRE) \cite{NRE}.  NRE, and the other methods we consider, use a neural network to approximate functions of the likelihood and/or posterior.  With the right choice of loss function, the neural network approximations will converge to the exact function being approximated during training on many simulated realizations of the data.  The resultant posterior estimates are sufficiently accurate for most purposes.

There are several advantages to our proposed SBI approach to CMB cluster lensing.  First, unlike the quadratic estimator approach, our SBI approach should naturally and close-to-optimally use all information in the data.  Indeed, we will show that our SBI method yields cluster mass constraints that are in agreement with the exact (and optimal) likelihood calculation in scenarios where the likelihood can be easily computed.  This presents a significant advantage over quadratic estimator approaches.  Second, SBI allows for the handling of non-idealities in the data (e.g. lensing from multiple structures, telescope beam and filtering) that cannot be easily incorporated into the exact likelihood.  Third, our SBI analysis directly produces constraints on cluster masses, without requiring intermediate steps like estimating the lensing convergence or measuring a lensing profile and estimating its covariance.  An additional advantage of the SBI approach introduced here is its simplicity: once the simulator has been created, training the neural likelihood/posterior models is easy, fast and robust to details of implementation (as we demonstrate below).  Once the posterior model is trained, it is very fast (essentially instantaneous) to compute constraints on cluster masses from new data realizations.  This last property may become relevant for future large samples of clusters.

We also note that while the exact likelihood, MAP, and quadratic estimator approaches may be appealing because of their analytical nature, these methods often rely implicitly on simulations.  For instance, \cite{Baxter:2015} and \cite{Raghunathan:2017} used a hybrid simulation and exact likelihood approach, where simulations were used to incorporate the impact of the beam into the likelihood calculation, the MAP method of \cite{Saha:2023} uses simulations for calculation of the normalization, and the method of \cite{Raghunathan:2019} relies on simulations for calibration of the model for the lensing dipole.  

The outline of the paper is as follows.  In \S\ref{sec:methods} we describe our methodology, including the calculation of the exact likelihood in an idealized scenario, and the SBI method.  In \S\ref{sec:results} we present our main results, including demonstrating that the SBI method recovers the exact likelihood in the limit that the likelihood can be easily computed, and that it continues to perform well outside of this limit.  In \S\ref{sec:sbi_robust} we test the robustness of our results to variation in our SBI method. 
We conclude in \S\ref{sec:discussion}.  

Throughout, we assume a flat $\Lambda$CDM cosmological model with parameters consistent with the results of the \textit{Planck} 2018 analysis \citep{Planck18}

\section{Methodology}
\label{sec:methods}

\subsection{Exact likelihood in an idealized scenario}
\label{sec:exact_likelihood}

If we consider CMB cluster lensing to be a single, parameterizable remapping of a Gaussian random field and we ignore the impact of the telescope beam and filtering, then the exact likelihood for the CMB observations can be easily computed, as we now describe.
For a Gaussian random field with power spectrum $C_{\ell}$, the pixel-pixel covariance matrix is
\begin{equation}
\label{eq:unl_cov}
    \mathbf{C}{ij} = g(|\vec{\theta}_{ij}|) \equiv \sum_{l} C_{\ell} \frac{2l+1}{4\pi} J_0(l|\vec{\theta}_{ij}|),
\end{equation}
where $J_0$ is the zeroth order Bessel function of the first kind and $|\vec{\theta}_{ij}|$ is the magnitude of the angular separation between pixels $i$ and $j$, and vectors are in the image plane.   We have assumed small angles, as appropriate for the cluster lensing context.  Below, we will assume square pixel grids of dimension $N_{\rm pix} \times N_{\rm pix}$, where each pixel is $\theta_{\rm pix} = 0.5'$ on a side and we will explore different choices of $N_{\rm pix}$.   In the idealized scenario considered in this subsection, we will assume that all large scale structure (LSS) along the line of sight to the cluster is behind (i.e. at higher redshift than) the cluster.   In practice, this is not a bad assumption since CMB lensing is most sensitive to structure at $z \sim 2$, and the clusters of interest are typically at $z_c < 2$, so most lensing by LSS does occur at higher redshift than the cluster.  We will also assume in this idealized scenario that even after lensing by LSS, the CMB remains Gaussian, which is not exactly correct.   We compute the LSS-lensed CMB power spectrum using \texttt{CAMB}\footnote{\url{camb.info}} \citep{CAMB1, CAMB2}.  

We write the angular deflection field caused by cluster lensing as $\vec{\alpha}_i(M)$, where $i$ indexes the pixel and we use $M$ to generically represent some parameter(s) describing the mass distribution of the cluster.  For instance, this could be the total mass of the cluster or a parameter describing the shape of the mass profile.
The deflection field is a continuous vector field, but assuming the pixel size is sufficiently small compared to fluctuations in the map, little information will be lost by pixelizing.

Since the covariance of two pixels in the unlensed CMB depends only on the magnitude of the angular separation between the pixels, the crucial quantity for describing the lensed CMB is the \textit{difference} in deflection angles for the two pixels: if two pixels are deflected by the same amount, then their angular separation would remain unchanged, and their covariance before and after lensing would be the same.  We write the difference in deflection angles for two pixels as 
\begin{eqnarray}
    \vec{\delta}_{ij}(M) = \vec{\alpha}_i(M) - \vec{\alpha}_j(M).
\end{eqnarray}
Given a mass model for the cluster, $\vec{\alpha}_i(M)$ and thus $\vec{\delta}_{ij}(M)$ can be computed, as we discuss in \S\ref{sec:massmodel}.

The likelihood for the observed CMB  in the simplified scenario considered in this section is then
\begin{equation}
\label{eq:simple_likelihood}
    P(\vec{d}_{\rm obs} | M) = \mathcal{N}(\vec{d}_{\rm obs} ; \vec{\mu} = 0, \mathbf{C} = \mathbf{C}'(M) + \mathbf{N}),
\end{equation}
where $\mathbf{C}'(M)$ is the covariance of the lensed CMB field, given by
\begin{equation}
    \mathbf{C}'_{ij}(M) = g(|\vec{\theta}_{ij} + \vec{\delta}_{ij}(M)|),
\end{equation}
where $g(\theta)$ is given by Eq.~\ref{eq:unl_cov} with $C_l$ set equal to the LSS-lensed CMB power spectrum, and where $\mathbf{N}$ is the noise covariance matrix of the observations.  For many CMB observations, it is reasonable to assume that the noise covariance matrix is proportional to the identity matrix:
 \begin{eqnarray}
     \mathbf{N} = \sigma_N^2 \mathbb{I}_{N_{\rm pix}},
 \end{eqnarray}
 where $\sigma_{N}^2$ is set by the sensitivity of the observations.  We will set $\sigma_N^2 = (1\,\mu{\rm K \,arcmin})^2/\theta_{\rm pix}^2$.  This is roughly the expected noise level for CMB-S4 \citep{S4}, and comparable to the noise levels of existing deep-field observations from SPT-3G \citep{3G}, and planned observations from the upcoming Simons Observatory \cite{SO}.  A non-white noise spectrum could be easily incorporated with appropriate modification to the noise covariance.

\subsection{Modeling the cluster lensing deflection field}
\label{sec:massmodel}

In order to model $\vec{\alpha}(M)$, we begin by modeling the cluster mass distribution.  On average, the density profiles of clusters, $\rho(r)$, are well-described by spherically symmetric Navarro-Frenk-White (NFW) profiles \cite{NFW}, given by 
\begin{eqnarray}
    \rho(r) = \frac{\rho_0}{(r/r_s)(1+r/r_s)^2},
\end{eqnarray}
where $\rho_0$ and $r_s$ are parameters of the model.  We adopt the spherical overdensity mass convention, such that the cluster mass is
\begin{eqnarray}
    M_{\Delta c} = \Delta (4\pi/3) R_{\Delta c}^3 \rho_{\rm crit}(z_c),
\end{eqnarray}
where $\rho_{\rm crit}(z_c)$ is the critical density of the Universe at the redshift of the cluster, $z_c$ and $R_{\Delta}$ is the cluster-centric radius at which the mean enclosed density is $\Delta \rho_{\rm crit}(z_c)$.  The concentration parameter is defined by 
\begin{eqnarray}
    c_{\Delta c} = R_{\Delta c}/r_s.
\end{eqnarray}
Our convention will be to take $\Delta = 200$.  The main mass parameter of our analysis is then $M_{200c}$, while $c_{200c}$ impacts the shape of the profile.  We fix $c_{200c}$ using the mass-concentration relation from \cite{Diemer:2019}.  At around $R_{200c}$, the density profile will begin to depart from the NFW form owing to the presence of correlated LSS, i.e. the two-halo term \citep{CooraySheth}.  For simplicity, we ignore this contribution here, since most of the information about the cluster mass comes from near the cluster center.   We postpone a full treatment of the two-halo term to future work.

The lensing convergence, $\kappa$, is related to the line of sight integral of the cluster mass distribution:
\begin{eqnarray}
\kappa(\vec{\theta}) = \frac{1}{\Sigma_{\rm crit}}\int dl \, \rho(r(\vec{\theta}, l)),
\end{eqnarray}
where $l$ is the line of sight distance, $\vec{\theta}$ is a vector in the image plane, and $r(l,\vec{\theta})$ is the cluster-centric distance corresponding to a point at $l$ and $\vec{\theta}$.  The critical surface density, $\Sigma_{\rm crit}$,
is given by
\begin{eqnarray}
    \Sigma_{\rm crit} = \frac{c^2 d_s }{4\pi G d_l d_{ls}},
\end{eqnarray}
where $d_l$ and $d_s$ are the angular diameter distances to the lens (cluster) and source (CMB), respectively, and $d_{ls}$ is angular diameter distance between these redshifts.  In practice, we use \texttt{Colossus} \citep{Colossus} to compute the lensing convergence profiles of the clusters.

The lensing deflection field, $\vec{\alpha}$, is related to the convergence field via
\begin{eqnarray}
\label{eq:convergence_deflection}
\kappa = \frac{1}{2} \nabla \cdot \vec{\alpha}.
\end{eqnarray}
We convert between deflections and the lensing convergence in Fourier space through the use of FFTs.  We now have all of the ingredients needed to compute the likelihood from Eq.~\ref{eq:simple_likelihood}.

\subsection{Modeling lensing by uncorrelated LSS}
\label{sec:lss}

We now move beyond the simple treatment of \S\ref{sec:exact_likelihood}, where all lensing by LSS was treated as occuring at redshifts greater than that of the cluster.  Far away from the cluster, the LSS can be treated as effectively uncorrelated with the cluster (as noted previously, we ignore nearby correlated structure).  We separate these uncorrelated contributions to the total LSS lensing signal into two parts, one  from $z > z_c$ and one from $z < z_c$.  We compute the power spectra of these two contributions using the Limber approximation \cite{Limber:1953, Loverde:2008}:
\begin{eqnarray}
    C_{l}^{\kappa\kappa}(z_{\rm min}, z_{\rm max}) = \int_{\chi(z_{\rm min})}^{\chi(z_{\rm max})} d \chi \frac{W^2(\chi)}{\chi^2} P_{\rm NL}\left(\frac{l+1/2}{\chi}, z(\chi)\right),
\end{eqnarray}
where $\chi(z)$ and $z(\chi)$ are the comoving distance and redshift to redshift $z$ and comoving distance $\chi$, respectively, $P_{\rm NL}(k,z)$ is the nonlinear matter power spectrum at wavenumber $k$ and redshift $z$, which we compute using \texttt{CAMB} with the \texttt{halofit} \cite{Takahashi:2012} prescription for nonlinear evolution, and 
\begin{equation}
    W(\chi) =  \frac{3\Omega_m H_0^2\chi}{2c^2a(\chi)} \frac{\chi^* - \chi}{\chi^*},
\end{equation}
 is the lensing weight function, where $a(\chi)$ is the scale factor corresponding to comoving distance $\chi$,  and $\chi^*$ is the comoving distance to the last scattering surface.  If we consider structure at redshifts below that of the cluster, then $(z_{\rm min}, z_{\rm max}) = (0, z_c)$, while if we consider structure at redshifts above the cluster, then $(z_{\rm min}, z_{\rm max}) = (z_c, z(\chi^*))$.

\subsection{Astrophysical contaminants}

A number of additional sources can contribute to CMB observations near galaxy clusters besides the lensed CMB and noise.  These include, for example, the thermal (tSZ) and kinematic (kSZ) SZ effects, dusty galaxies and radio galaxies.  Perhaps the most significant of these is the thermal SZ signal, which can significantly bias the inference of CMB cluster lensing \cite[e.g.][]{Baxter:2015, Baxter:2018}. In principle, there are several techniques one could use to mitigate the impact of these sources.  To remove the tSZ one could combine CMB maps at multiple frequencies to null the frequency-dependent tSZ signal, although this will come at the cost of an increase in noise levels.  Alternatively, one could use polarization data instead of temperature data to extract the CMB cluster lensing signal.  Astrophysical sources typically produce much smaller biases to lensing estimation from CMB polarization than from CMB temperature.  However, using CMB polarization instead of temperature will also lead to some reduction in signal-to-noise.  Here, for simplicity, we ignore the presence of the additional astrophysical sources.  In principle,  using tSZ-cleaned maps or polarization data instead of temperature data would not significantly change the SBI methodology that we introduce.  We postpone a more detailed treatment of astrophysical foregrounds to future work.

\subsection{Simulator}

Our SBI analysis relies on having the ability to generate mock realizations of the cluster-lensed  CMB given a choice of parameters for the cluster mass distribution.  We generate simulated cluster-lensed CMB maps with the following steps:

\begin{enumerate}
\item \textbf{Draw an unlensed CMB realization}: $\vec{d}_{\rm unl} \sim \mathcal{N}(\vec{\mu} = 0, \mathbf{C})$, where $\mathbf{C}$ is given by Eq.~\ref{eq:unl_cov}.

\item \textbf{Compute the convergence field due to the cluster} at parameter values $M$ following the steps in \S\ref{sec:massmodel}

\item  \textbf{Generate Gaussian random convergence fields from LSS} at $z > z_c$ and $z < z_c$ using the power spectra described in \S\ref{sec:lss}.

\item \textbf{Apply the successive lensing operations to the unlensed CMB realization}.  Convert the three convergence fields into deflection fields using Eq.~\ref{eq:convergence_deflection}, and apply the deflections in succession to the unlensed CMB realization from (1).  That is, first lens by the LSS at redshifts above the cluster, then lens the resultant map by the cluster, then lens the result by LSS at redshifts below the cluster.  For each lensing operation, we use fifth order spline interpolation to interpolate the unlensed map at the positions corresponding to the deflected pixels.
 
 \item \textbf{Apply beam}.  We apply a Gaussian beam with $\sigma = 1'$ to the lensed CMB maps.  This beam size is comparable to that of SPT and CMB-S4.  

 \item \textbf{Add instrumental noise}. Draw a Gaussian realization from the noise covariance matrix $\mathbf{N}$ and add it to the map. 
\end{enumerate}

We use pixels of size $0.5'$ throughout.  The convergence maps are computed using $N_{\rm pix} = 32$, the unlensed CMB maps are computed using $N_{\rm pix}$ = 16, and the final SBI analysis is restricted to a region with $N_{\rm pix} =  8$ centered on the cluster.

\subsection{Neural likelihood model}

We now describe the SBI methodology that we apply to estimate cluster masses from the CMB lensing signal. As noted above, the use of SBI is motivated by the intractability of the exact likelihood in the general problem, while simulating samples from the forward model is easy.  

Our baseline analysis uses the neural likelihood ratio estimator (NRE) method of \cite{NRE}. This method trains a neural classifier to learn the likelihood ratio between different points in parameter space, given the data.  The estimator is amortized in the sense that it is trained to work for different possible realizations of the data, not just the actual observed data.  Briefly, given a set of observations $x$ and parameters $\theta$, the NRE method trains a classifier to distinguish between samples $(x,\theta)$ obtained from two different probability distributions. First is the joint parameter distribution, $p(x,\theta)$ and second is the product of the marginal distributions $p(x)p(\theta)$. We use the lensed CMB maps as our observations $x$ and the mass of the cluster as the parameter $\theta$. We use a residual network to extract the features from the data and train the neural network as a classifier to learn the optimal decision function distinguishing samples obtained from $p(x,\theta)$ versus from $p(x)p(\theta)$ which is given by:
\begin{equation}
s(x,\theta) = \frac{p(x,\theta)}{p(x,\theta) + p(x)p(\theta)}.
\end{equation}

The decision function $s(x,\theta)$ is directly related to the likelihood ratio $r = p(x,\theta)/(p(x) p(\theta))$ through $r = s/(1-s)$. Finally, once the likelihood ratio estimator is trained, standard MCMC can be used to generate samples from the posterior.  We use the default NRE architecture as implemented in the \texttt{sbi} package \cite{sbicode}. We use 90000 simulations for the training set and 10000 simulations as validation set.

Note that there exist other neural network-based SBI approaches to estimate posteriors such as neural posterior estimate (NPE; \cite{NPE}) and neural likelihood estimate (NLE; \cite{SNL}). 
The NPE and NLE estimator aims to learn the posterior, $p(\theta|x)$, or the likelihood, $p(x|\theta)$, by modeling this conditioned probability distributions using stacks of normalizing flows.  NLE is generally followed by a MCMC step to obtain the posterior from the learned likelihood. In this study we use the stacks of masked autoregressive flows in both NLE and NPE methods, again following the default architecture of the \texttt{sbi} package \cite{sbicode}.

We perform a study of robustness of our results to these different approaches in \S\ref{sec:sbi_robust}.

\subsection{Stacking posteriors}

The constraints on $M$ from a single cluster can be very weak, even when the instrumental noise is very low.  This is because variance in the unlensed CMB contributes significantly to variance in the signal.    Without a significant gradient in the unlensed CMB behind the cluster, there will be no cluster lensing signal \citep{Seljak:2000}.  Because the constraints from individual clusters can be weak, it is standard practice to enhance the signal-to-noise of the CMB cluster lensing measurements by combining mass constraints  across multiple clusters (i.e. stacking).   In the quadratic estimator approach to CMB cluster lensing,  stacking is often done at the level of the convergence maps.  Stacking the convergence maps with equal weighting (as is often done) can be suboptimal, since the signal-to-noise for each cluster varies widely due to the different background CMB realizations. 
 Here,  we instead ``stack'' the likelihoods (or the SBI approximation to the likelihoods), by taking the product across the individual cluster likelihoods.  Each cluster can be assumed to be independent, since they are likely to be well separated on the sky.  Given our assumptions, stacking the likelihoods will yield optimal constraints on the average cluster mass of the stacked clusters. This is  essentially the  quantity of interest for cosmological analyses since it sets the normalization of the mass-observable relation.  We note that the NFW model for the cluster mass distribution may not describe individual clusters very well.  For instance, individual clusters may be triaxial \cite{Jing:2002}.  By computing the likelihood for individual clusters and then stacking the resultant likelihoods, we may incur some bias from departures from the NFW form.  In principle, this bias could be ameliorated with a more flexible mass model.  We postpone a detailed investigation into departures from our assumed NFW model to future work.

\section{Results}

\label{sec:results}

\begin{figure}
    \centering
    \includegraphics[scale=0.5]{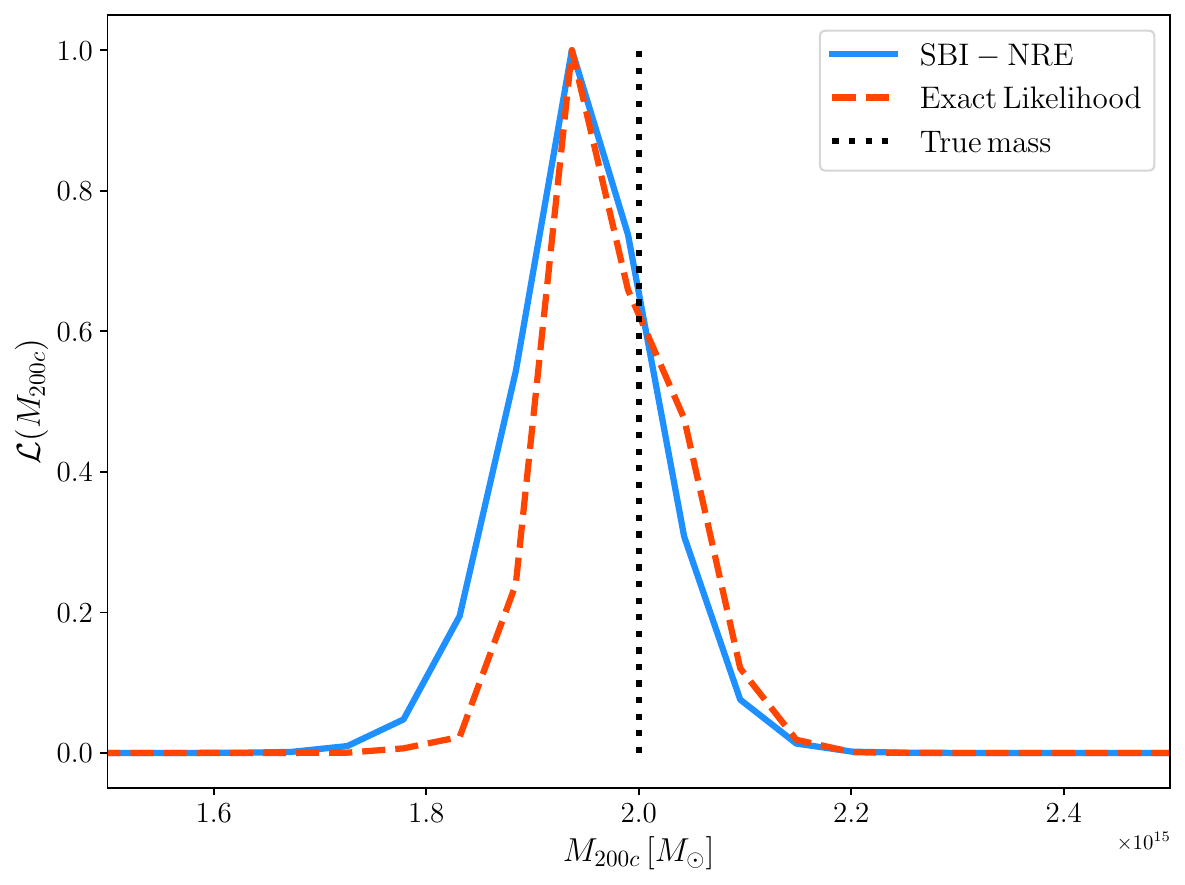}
    \includegraphics[scale=0.5]{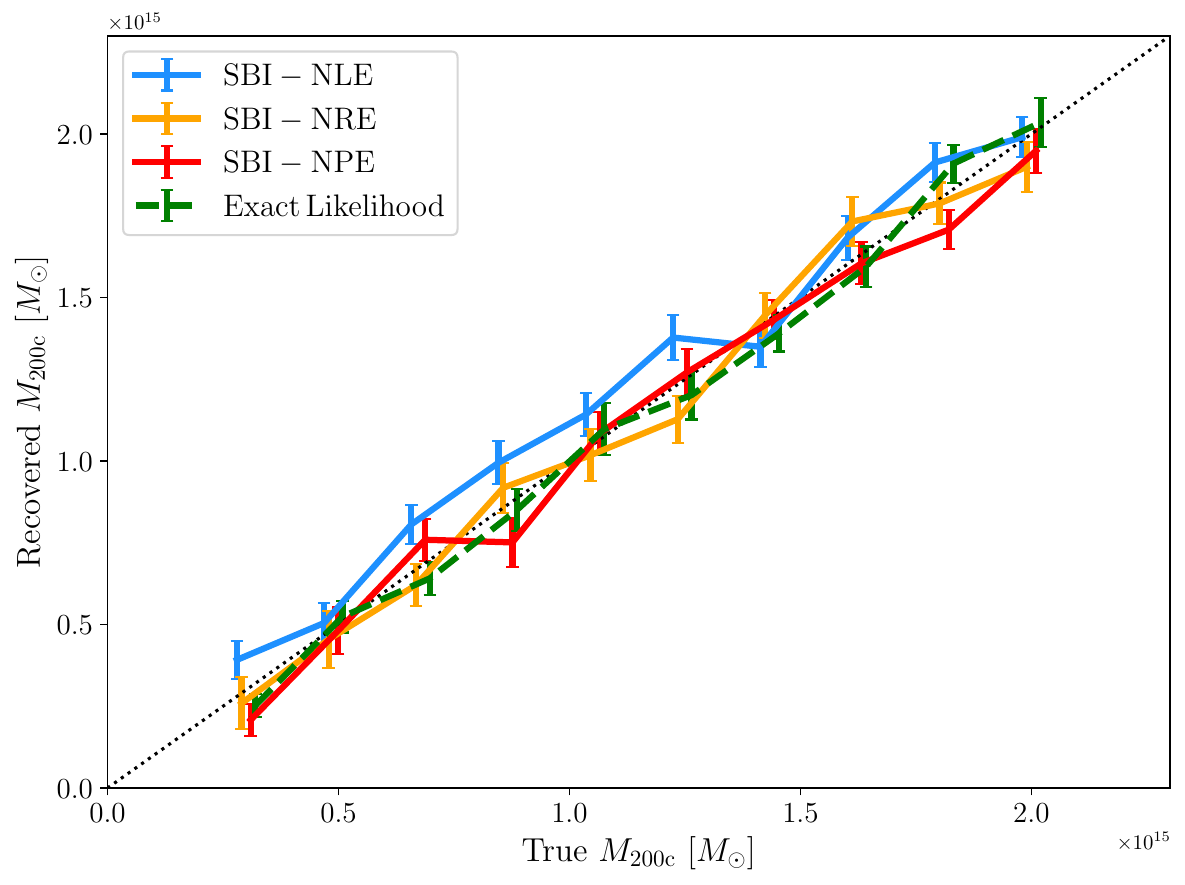}
    \caption{Comparison of the stacked exact likelihood and the SBI results for the idealized case where the exact likelihood can be computed.  Top: the blue (SBI) and red (exact likelihood) curves represent combined constraints from 20 clusters, where the true mass is indicated by the vertical dotted line. 
 The SBI estimate of the stacked likelihood is close to that from the exact calculation.  Bottom: We show the recovered mean and standard deviation of the exact likelihood (red) and SBI (blue) constraints on the cluster mass as a function of the true cluster mass.  Each point represents a stack over 20 clusters with the same mass.  We see that the SBI constraints are consistent with those of the exact likelihood over a range of cluster masses. }
    \label{fig:sbi_vs_likelihood_simple}
\end{figure}

\subsection{Comparison to the exact likelihood in idealized scenario}
\label{sec:lensing_ideal}
 We first apply our SBI analysis in the simplified case where the exact likelihood can be quickly computed (i.e. no instrumental beam, and treating lensing as a single remapping of a Gaussian random field, as described in \S\ref{sec:exact_likelihood}).    We generate a set of 20 mock lensed CMB observations using this model.  The cluster mass and redshift for this set of simulations is the same for each cluster: $M_{200c} = 3\times 10^{15}\,M_{\odot}$ and $z_c = 0.5$.  Fig.~\ref{fig:sbi_vs_likelihood_simple} shows the resultant stacked likelihoods from these 20 clusters (top panel) as a function of $M_{200c}$. The input cluster mass is indicated with the vertical dotted line.  As expected the exact likelihood peaks near to the input cluster mass.  The SBI analysis recovers a posterior consistent with the exact likelihood.  In the bottom panel, we repeat this test across a range of cluster masses.  We find that the posteriors from SBI agree very well with those of the exact likelihood across the range of cluster masses considered.  These results suggest that the SBI method is close-to-optimally extracting the information in the data about the cluster mass.

\subsection{Comparison to exact likelihood in general scenario}
\label{sec:lensing_general}
We next consider the case where the data come from the more general model that includes the instrumental beam and the more correct lensing implementation, with high-$z$ and low $z$ structure are treated separately.  We train the SBI model on data of this type, but preserve the same simple  likelihood calculation for comparison.  

\begin{figure}
    \centering
    \includegraphics[scale = 0.5]{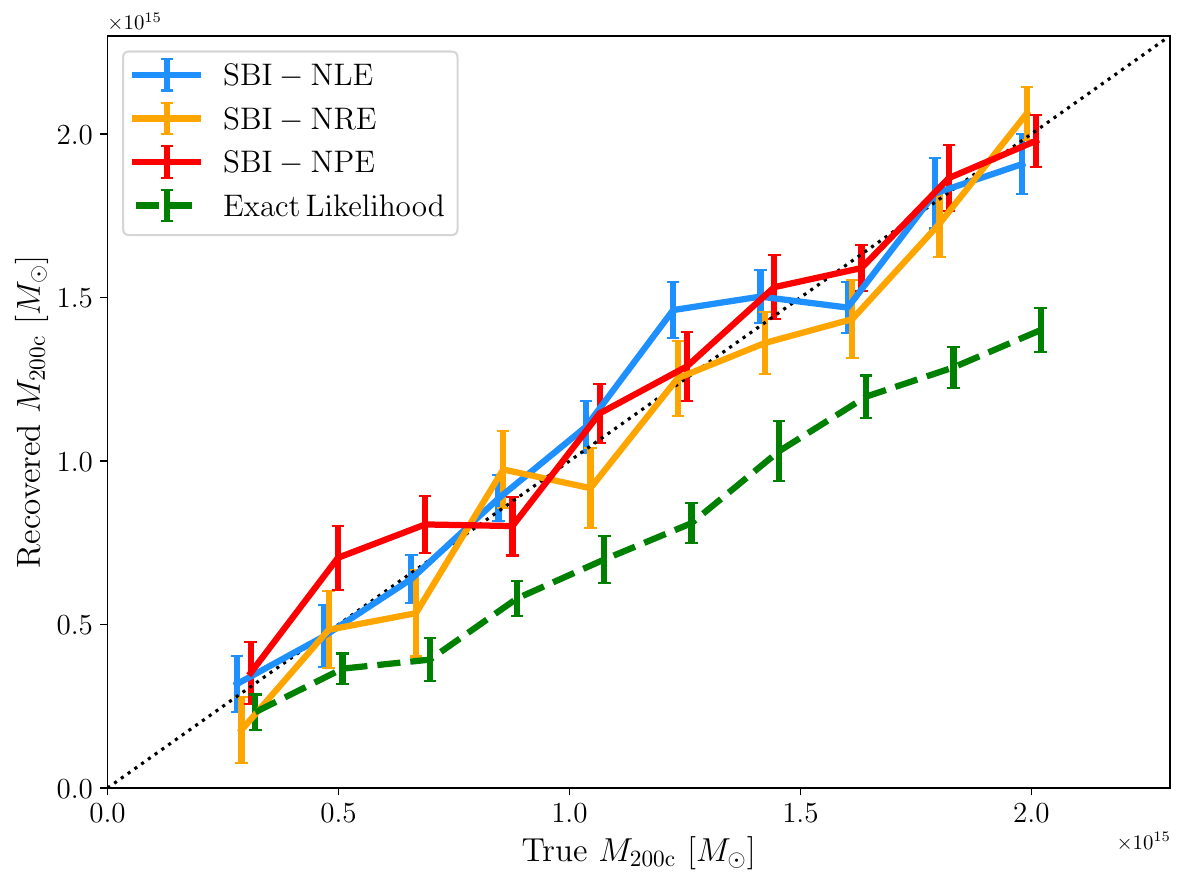}
    \caption{Same as bottom panel of Fig.~\ref{fig:sbi_vs_likelihood_simple}, except now the data are generated using the full lensing model and include an instrumental beam.  In this case, the SBI analysis continues to recover good estimates of the cluster mass.  The simple likelihood calculation (green dashed), on the other hand, is significantly biased.  The solid curves indicate the results when we vary the details of the SBI methodology.}
    \label{fig:sbi_vs_likelihood_general}
\end{figure}

The results of this test are shown in Fig.~\ref{fig:sbi_vs_likelihood_general}.  As expected, SBI continues to recover reasonable estimates of the true cluster masses.  The constraints on the cluster masses are unbiased and have comparable uncertainty to the case where the beam and other effects are not included.  We would expect the presence of the beam to slightly degrade the cluster mass constraints since a 1 arcmin beam is comparable to the scale of the lensing defelections.  Indeed, the SBI constraints become somewhat weaker in the presence of the beam.  The results of the likelihood calculation that ignores the instrumental beam and the different impact of high and low-$z$ lensing  are shown with the red errorbars.  We see that the beam and LSS lensing significantly bias the simple likelihood calculation.  Essentially all of the bias comes from ignoring the beam, while the treatment of LSS has little impact on the constraints.  Apparently, the likelihood calculation is very sensitive to modeling the beam correctly.  It is not too surprising that the treatment of LSS has a small impact on the results, since effectively all that matters for the cluster lensing signal is the local gradient of the unlensed CMB, and the typical size of this gradient will not be significantly changed by LSS lensing, which mostly modifies the small-scale CMB power spectrum.

\section{Robustness to variations in SBI methodology}
\label{sec:sbi_robust}

We now explore the sensitivity of our results to different network architectures and SBI methods by obtaining posteriors with NLE, NRE and NPE. We unify the training and test datasets used in all three methodologies. The results are shown  with the different solid curves in Fig.~\ref{fig:sbi_vs_likelihood_general}.  We see that all three SBI methods give consistent posteriors, suggesting that the SBI analysis is highly robust to neural network implementation. 

\section{Discussion}
\label{sec:discussion}

We have presented the first application of SBI to the problem of estimating cluster masses from gravitational lensing of the CMB, and to our knowledge, the first application of NRE to CMB lensing.  The CMB cluster lensing problem is well-suited to SBI since it is relatively easy to simulate cluster-lensed CMB maps, but dificult to compute their likelihood in full generality.  We find that in idealized cases where the exact likelihood can be computed, our SBI method recovers the exact likelihood well, indicating that this method is close-to-optimally exracting the cluster mass information from the data.  In the general problem for which the likelihood cannot be easily computed, SBI recovers unbiased estimates of the cluster mass.  These constraints are quite robust to the details of the SBI implementation.  SBI with neural posterior/likelihood models therefore present a powerful set of tools for extracting cluster mass estimates from current and future observations of the lensed CMB.  

We plan to make the code used for this analysis publicly available at \url{https://github.com/ebaxter/CMBClusterLens} upon acceptance of the paper by a journal.

\acknowledgments

We thank Peter Sadowski for useful discussions related to this work, and Srinivasan Raghunathan for detailed comments. SP is supported by the Simons Collaboration on Learning the Universe.  EB is partially supported by NSF Grant AST-2306165.

\bibliographystyle{unsrt}
\bibliography{thebib.bib}

\end{document}